\documentclass[a4paper]{jpconf}
\usepackage{graphicx}
\usepackage{amssymb}
\usepackage{mathrsfs} 
\begin{document}
\def\d{{\mathrm{d}}}
\newcommand{\scri}{\mathscr{I}}
\newcommand{\sun}{\ensuremath{\odot}}
\def\J{{\mathscr{J}}}
\def\sech{{\mathrm{sech}}}
\title{Entropy bounds for uncollapsed matter}

\author{Gabriel Abreu and Matt Visser}

\address{School of Mathematics, Statistics and Operation Research\\ 
Victoria University of Wellington\\ 
Wellington, New Zealand.}

\ead{Gabriel.Abreu@msor.vuw.ac.nz; Matt.Visser@msor.vuw.ac.nz}

\begin{abstract}
In any static spacetime the quasilocal Tolman mass contained within a volume can be reduced to a 
Gauss-like surface integral involving the flux of a suitably defined generalized surface gravity. By introducing some basic thermodynamics, and invoking the Unruh effect, one can then develop elementary bounds on the quasilocal entropy that are very similar in spirit to the holographic bound, and closely 
related to entanglement entropy.

\end{abstract}

\section{Tolman mass and generalized surface gravity}
Among the different notions of quasi-local mass, the Tolman mass has the advantage that when dealing with a horizonless object (like a star/planet/monster/gravastar/black star/quasi-black hole) in a static spacetime, it can ultimately be bounded in terms of quantities measurable in the surface of the object \cite{Abreu:2010sc}; particularly the average surface gravity and the total area. In any static spacetime where  the metric can be put into the form
\begin{equation}
\d s^2=-e^{-2\Psi}\d t^2+g_{ij}\d x^i\d x^j,
\end{equation}
 the Tolman mass \cite{TolmanM} inside a region $\Omega$ is defined in terms of the diagonal components of the stress-energy tensor as follows:
\begin{equation}
m_T(\Omega)=\int_\Omega\sqrt{-g_4}\left(T_0^0-T_i^i \right)\,\d^3x.
\end{equation}
Here $g_4$ is the determinant of the $(3+1)$-dimensional metric. The Einstein equations then imply
\begin{equation}
m_T(\Omega)=\frac{1}{4\pi}\int_\Omega\sqrt{-g_4}\,R^0_0\,\d^3x,
\end{equation}
which is a purely geometrical statement. Now, it is possible to rewrite the Tolman mass in terms of the three-acceleration of a specific set of fiducial observers (FIDOs). First consider the set of FIDOs with four-velocity parallel to the time-like Killing vector $k^a$,
\begin{equation}
V^a= {k^a\over ||k||} = \left(\sqrt{|g^{00}|},\,0,\,0,\,0 \right).
\end{equation}
Their four-acceleration is given by
\begin{eqnarray}
\label{4A1}
A^a=V^b\nabla_bV^a=V^0\left(\partial_0V^a+\Gamma^a{}_{c0}V^c \right)= \sqrt{|g^{00}|}\,\Gamma^a{}_{00}\sqrt{|g^{00}|}=|g^{00}|\,\Gamma^a{}_{00}.
\end{eqnarray}
That is, $A^a=\left(0;\,|g^{00}|\,\Gamma^i{}_{00}\right)$, since $V$ is four-orthogonal to $A$. 

On the other hand, consider the Raychaudhuri equation for this timelike congruence:
\begin{equation}
\frac{\d\theta}{\d s}=-R_{ab}\,V^a\,V^b+2\,\omega^2-2\,\sigma^2-\frac{1}{3}\theta^2+\nabla_a A^a.
\end{equation}
Here the expansion $\theta$, shear expansion $\sigma_{ab}$, and vorticity $\omega_{ab}$ of the congruence are defined as usual. In terms of the spatial projection operator $h_{ab}=g_{ab}+V_a\,V_b$ we have:
\begin{eqnarray}
\theta_{ab}=h_{ac}\,\nabla^{(c}V^{d)}\,h_{db}&;&\qquad  \theta=g^{ab}\,\theta_{ab}=\nabla_a\,V^a;\nonumber \\
\sigma_{ab}=\theta_{ab}-\frac{1}{3}\,h_{ab}\,\theta&;&\qquad\sigma^2=\frac{1}{2}\,\sigma_{ab}\sigma^{ab};\nonumber\\
\qquad \omega_{ab}=h_{ac}\nabla^{[c}\,V^{d]}\,h_{db}&; &\qquad \omega^2= {1\over2}\,\omega_{ab}\,\omega^{ab}.
\end{eqnarray}
Then, since $k_{(a;b)}=0$ in view of Killing's equation, we have
\begin{equation}
V_{(a;b)}=\nabla_{(a}\left\{k/||k|| \right\}_{b)}=\frac{V_{(a}\,||k||_{,b)}}{||k||}.
\end{equation}
Therefore  $\theta_{ab}=0$, and consequently $\theta=0$ and $\sigma_{ab}=0$. Furthermore, we also have $V_a = -|g_{00}|^{-1/2} \; (1,0,0,0)_a$, that is,  in the notation of differential forms $V^\flat = e^\Psi \; \d t $. This implies $\d V^\flat = \d\Psi \wedge V^\flat$ and consequently $\omega_{ab}=0$ (the congruence of FIDOs is hypersurface orthogonal), implying $\omega=0$. 
Thus in this situation the timelike Raychaudhuri equation implies
\begin{equation}
R_{ab}\,V^a\,V^b=\nabla_a A^a.
\end{equation}
This finally reduces to the well known Landau--Lifshitz result~\cite{LL},
\begin{equation}
R_0^0=
\frac{1}{\sqrt{-g_4}}\partial_i\left(\sqrt{-g_4}\,A^i \right).
\end{equation}
Now using ordinary partial derivative integration by parts, for any $3$--volume $\Omega$ the Tolman mass can be rewritten as
\begin{eqnarray}
m_T(\Omega)&=&\frac{1}{4\pi}\int_\Omega \partial_i\left(\sqrt{-g_4}\,A^i \right)\,\d^3x,\nonumber\\
&=&\frac{1}{4\pi}\int_\Omega\partial_i\left(\sqrt{g_3}\,e^{-\Psi}\,A^i\right)\,\d^3x,\nonumber\\
&=&\frac{1}{4\pi}\int_{\partial\Omega}\left(e^{-\Psi}\,A^i \right)\,\hat{n}_i\,\sqrt{g_2}\,\d^2x,
\end{eqnarray}
where $\hat{n}$ is the unit normal, and $\sqrt{g_2}$ is the induced area measure on $\partial\Omega$.

We now consider a  generalization of the surface gravity. One method to obtain the usual static surface gravity is by considering the magnitude of the four-acceleration of a suitable set of FIDOs, and renormalizing it using a suitable redshift factor. Here, instead we take the three spatial components of such four-acceleration to build a three-vector
\begin{equation}
\kappa^i=e^{-\Psi}\,A^i,
\end{equation}
with magnitude $\kappa=e^{-\Psi}\sqrt{g_{ij}A^i\,A^j}$. This generalized surface gravity retains the \emph{usual} physical interpretation of the surface gravity, but  is not only valid for any event horizon present --- this definition now also applies to FIDOs on the (arbitrary) boundary $\partial\Omega$. The Tolman mass is
\begin{equation}
m_T(\Omega)=\frac{1}{4\pi}\int_{\partial\Omega}\kappa^i\,\hat{n}_i\,\sqrt{g_2}\,\d^2x.
\end{equation}
Furthermore, defining an average surface gravity over the boundary $\partial\Omega$, 
\begin{equation}
\bar{\kappa}(\partial\Omega)\equiv\frac{\int_{\partial\Omega}\kappa\,\d(area)}{\int_{\partial\Omega}\,\d(area)}, 
\end{equation}
and total area $\mathscr{A}(\partial\Omega)$, and assuming there are no black holes regions, we finally have
\begin{equation}
m_T(\Omega)\leq\frac{\bar{\kappa}(\partial\Omega)\;\mathscr{A}(\partial\Omega)}{4\pi}.
\end{equation}
This is a bound on the total Tolman mass of a horizonless object (such as a star, or a planet, or a monster \cite{Sorkin:1981wd}, or a gravastar \cite{Mazur:2004fk,Visser:2003ge,Cattoen:2005he}, or a black star \cite{Barcelo:2007yk,Barcelo:2009zz,Visser:2009pw}, or a quasi--black hole \cite{Lemos:2009uk,Lemos:2008cv}) in a general static geometry, which only depends on quantities measurable on its surface.

\section{Entropy bounds and the Unruh effect}
To take thermodynamics into account, and calculate a bound on the entropy of these types of object, we now need to consider the Euler (Gibbs--Duhem) relation for the entropy density of the matter forming the horizonless object,
\begin{equation}
s=\frac{\rho+p-\mu\,n}{T},
\end{equation}
where (as usual) we have set $p=\frac{1}{3}tr\{T_{ij}\}$ and $\rho=T_{00}$. It is important to remember that  the whole discussion in centered in some collection of ideal particles making up an uncollapsed object like a star/planet/monster/gravastar/black star/quasi-black hole. The total entropy due to this kind of matter inside any specific $3$--volume is
\begin{eqnarray}
S(\Omega)&=&\int_\Omega\sqrt{g_3}\;\frac{\rho+p+\mu\,n}{T}\,\d^3x.
\end{eqnarray}
The Tolman \cite{TolmanM,TolmanEq}, (Tolman--Ehrenfest~\cite{T-Ehren}), and Tolman--Klein \cite{TolmanKlein} equilibrium conditions
\begin{equation}
T\,\sqrt{-g_{00}}=T_\infty ,\qquad \mu\,\sqrt{-g_{00}}=\mu_\infty,
\end{equation}
imply
\begin{equation}
S(\Omega)=\frac{1}{T_\infty}\int_\Omega\sqrt{-g_4}\left(\rho+p \right)\,\d^3x-\frac{\mu_\infty}{T_\infty}\,N.
\end{equation}
Here $N\equiv\int_\Omega\sqrt{g_3}\,n\,\d^3x$ is the total number of particles inside the $3$--volume. Thermodynamic stability requires the chemical potential to be $\mu\geq0$, then
\begin{eqnarray}
S(\Omega)\leq\frac{1}{T_\infty}\int_\Omega\sqrt{-g_4}\,\left(\rho + p \right)\,\d^3x\leq\frac{1}{T_\infty}\int_\Omega\sqrt{-g_4}\,\left(\rho + 3p \right)\,\d^3x.
\end{eqnarray}
where we have assumed $p\geq0$ throughout the interior. This last numerator is just equivalent to the definition of the Tolman mass. Hence
\begin{equation}
S(\Omega)\leq\frac{m_T(\Omega)}{T_\infty}.
\end{equation}
We see that the total entropy is bounded by the Tolman mass divided by the temperature as measured at infinity. Furthermore from the previous discussion this entropy bound can be rephrased in terms of the average generalized surface gravity $\bar{\kappa}(\partial\Omega)$, and the total area $\mathscr{A}(\partial\Omega)$ ,
\begin{equation}
S(\Omega)\leq\frac{1}{4\pi\,T_\infty}\; \bar{\kappa}(\partial\Omega)\;\mathscr{A}(\partial\Omega).
\end{equation}

Finally we invoke the only semiclassical notion we use throughout the calculations, to relate the surface gravity to a temperature, via the Unruh effect \cite{Unruh:1976db}. Due to the existence of the Unruh acceleration radiation phenomenon, it can be asserted that an observer on the boundary $\partial\Omega$ will measure a minimum local temperature of
\begin{equation}
T(x)\geq T_\mathrm{Unruh}(x)=\frac{||A(x)||}{2\pi},
\end{equation}
which when redshifted to infinity implies
\begin{equation}
T_\infty \geq \max_{x\in\partial\Omega}\left\{\frac{\kappa(x)}{2\pi} \right\}.
\end{equation}
So the equilibrium temperature of the object confined inside $\partial\Omega$ satisfies
\begin{equation}
T_\infty\geq\frac{\bar{\kappa}(\partial\Omega)}{2\pi}.
\end{equation}
Hence the entropy is bounded by  \emph{one half} of the total area $\mathscr{A}(\partial\Omega)$,
\begin{equation}
S(\Omega)\leq\frac{\mathscr{A}(\partial\Omega)}{2}.
\end{equation}

\section{Discussion}
We have bounded the entropy of an object without horizons, in static spacetime, by half of its total area. Although the construction of this entropy bound is very minimalist, it is closely related to the usual entropy bounds \cite{Bousso:2002ju, Bekenstein:1974ax, Bekenstein:1980jp, Padmanabhan:2003pk, Padmanabhan:2010xh, Srednicki:1993im}. Specifically, it resembles the holographic bound \cite{Bousso:2002ju}, which corresponds to $S(\Omega)\leq\frac{1}{4}\mathscr{A}(\partial\Omega)$. 

The fundamental reason for the coefficient mismatch between these two bounds is because we calculated the entropy of a uncollapsed distribution of matter, where the temperature has its normal interpretation as an intensive variable, and the Euler relation keeps its usual uncollapsed form. Further generalizations of this entropy bound, where a black hole region is taken into account, can be found in \cite{Abreu:2010sc}. Overall, the importance of this entropy bound, for an uncollapsed distribution of matter, resides in its simplicity and its robustness.

\section*{References}

\end{document}